\documentclass[3p,times,twocolumn]{elsarticle}

\usepackage{ecrc,multirow,array}

\volume{00}

\firstpage{1}

\journalname{Nuclear Physics B Proceedings Supplement}

\runauth{}

\jid{nuphbp}

\jnltitlelogo{Nuclear Physics B Proceedings Supplement}

\usepackage{amssymb}

\usepackage[figuresright]{rotating}

\begin{document}

\begin{frontmatter}



\dochead{\small{IFJPAN-IV-2013-1, IFIC/12-76, UAB-FT-727}}

\title{RChL currents in Tauola: implementation and fit parameters}


\author[olga2,olga1,zbys1]{O.Shekhovtsova\fnref{sp}}
\address[olga2]{NSC Kharkov Institute of Physics and Technologyâ, Kharkov UA-61108, Ukraine}
\address[olga1]{IFIC, Universitat de Val\`encia-CSIC,  Apt. Correus 22085, E-46071,  Val\`encia, Spain}

\address[zbys1]{Institute of Nuclear Physics, PAN,
        Krak\'ow, ul. Radzikowskiego 152, Poland}
\author[ian]{I. M. Nugent}
\address[ian]{RWTH Aachen University, III. Physikalisches Institut B, Aachen, Germany}
\author[tom]{T. Przedzinski}
\address[tom]{The Faculty of Physics, Astronomy and Applied Computer
Science, \\Jagellonian University, Reymonta 4, 30-059 Cracow, Poland}
\author[pablo]{P. Roig}
\address[pablo]{Grup de F\'{\i}sica Te\`orica, Institut de F\'{\i}sica d'Altes Energies, 
Universitat Aut\`onoma de Barcelona, E-08193 Bellaterra, Barcelona, Spain}
\author[zbys1,zbys2]{Z. Was} 
\address[zbys2]{CERN PH-TH, CH-1211 Geneva 23, Switzerland}
\fntext[sp]{Speaker}

\begin{abstract}

We present the results of a partial upgrade to the Monte Carlo event generator TAUOLA using 
Resonance Chiral Theory for the two and three meson final states. These modes account for 88\% 
of total hadronic width of the tau meson.  The first results of the model parameters 
have been obtained using Preliminary BaBar data for 3$\pi$ mode. 
\end{abstract}


\end{frontmatter}


\section{Introduction}
\label{sec:intro}
The tau lepton is the heaviest of the leptons, and provides a unique opportunity to study low energy QCD and the mechanism of hadronization.  
With the shut-down of both the B-factory collaborations, BaBar \cite{babar} and Belle \cite{belle}, 
this is a critical time to make the results from more than a decade of experimental research 
available in a useful manner to high energy physics community before the opportunity is lost. 
To accomplish this, interaction between experimentalist and theorists is required to determine 
the optimal way for comparing experimental data with the theoretical prediction.
  The approach used by this work is to compare theory to data by using a Monte Carlo event generator to simulate the theoretical models.

The Monte Carlo event generator TAUOLA is a long term project that started in the 90's with the publication \cite{Jadach:1993hs}. 
The generator simulates more than twenty modes, including both the leptonic and hadronic modes. 
Modeling the hadronic decay modes involves matrix elements that convey the hadronization of the
vector and axial-vector currents. At present there is no determination from first principles 
for those matrix elements since they involve strong interaction effects in the non-perturbative regime. 
Therefore, one has to rely on models that parameterize the form-factors originating from the hadronization.
The hadronic currents implemented in the generator TAUOLA \cite{Jadach:1993hs} were based on theoretical results presented in \cite{Kuhn}. 

In these models, the hadronic form-factors are written as a weighted sum of products of Breit-Wigner functions.  
This approximation, as it is demonstrated in \cite{GomezDumm:2003ku}, 
is not able to reproduce the next-to-leading-order $\chi$PT results.
Later the experimental collaborations, both Cleo \cite{cleo,Coan:2004ep} and Aleph \cite{aleph}, 
introduced improvements based on the result of their data analysis and in some cases spoiling the theoretical constraints 
\footnote{In fact, hadronic currents of Cleo and Aleph versions spoil some theoretical prediction. 
For example, to reproduce $\tau^-\to (K K\pi)^- \nu_\tau $ data the Cleo collaboration \cite{Coan:2004ep} 
introduced two ad-hoc parameters and, as a result, the Wess-Zumino part does not reproduce the QCD normalization.}.
As an alternative, it was proposed to apply the methods of the Resonance Chiral Theory \cite{rcht}. 
This approach is a better theoretically founded than the approach of weighted products of Breit-Wigner functions. 
However, it requires comparison both BaBar and Belle experimental data with the results from theoretical 
models which have to be implemented into TAUOLA. In the next sections, we  will present 
the result on the implementation the RChT hadronic currents for $\pi^-\pi^0$, $K^-K^0$, $(K\pi)^-$, $(\pi\pi\pi)^-$ 
and $(K K\pi)^-$ modes in TAUOLA,  for details see  \cite{Shekhovtsova:2012ra, web:RChL}.

\section{Hadronic currents for two and three meson decay modes}\label{sec:hadr}
For any two meson $\tau$ decay channel, the most general form of hadronic currents ($J^\mu$), which is compatible with Lorentz invariance, is written as  
\begin{eqnarray}
J^\mu & = & N \bigl[ (p_1 - p_2 - \frac{\Delta_{12}}{s}(p_1 +p_2))^\mu F^{V}(s) 
\bigr. \label{eq:twomes} \\
& + & \bigl.
\frac{\Delta_{12}}{s}((p_1 + p_2)^\mu F^{S}(s) \bigr], \nonumber
\end{eqnarray}
where $p_1$ and $p_2$ are the momenta of hadrons, $\Delta_{12} = m_1^2 -m_2^2$, $s = (p_1 +p_2)^2$.
For a final state of three pseudoscalars
is 
\begin{eqnarray}
&&\!\!\!\!\!\!\!\!\!\!\!\!\!J^\mu = N \bigl\{T^\mu_\nu \bigl[ c_1 (p_2-p_3)^\nu F_1  + c_2 (p_3-p_1)^\nu
 F_2 \bigr.\bigr. \label{eq:fiveF} \\
&& \!\!\!\!\!\!\!\!\!\!\!\!\!\!\!\!\!\!\!\!\!\!\!\!\bigl.\bigl. + c_3  (p_1-p_2)^\nu F_3 \bigr] + c_4  q^\mu F_4  -{ i \over 4 \pi^2 F^2} c_5
\epsilon^\mu_{.\ \nu\rho\sigma} p_1^\nu p_2^\rho p_3^\sigma F_5      \bigr\},
\nonumber
\end{eqnarray}
where $p_1$, $p_2$ and $p_3$ are the momenta,
  $T_{\mu\nu} = g_{\mu\nu} - q_\mu q_\nu/q^2$ denotes the transverse
projector, $q^\mu=(p_1+p_2+p_3)^\mu$ is the momentum of the hadronic system   and
$F$ is the pion decay constant in the chiral limit. 
Among the form factors  $F_1$, $F_2$, $F_3$, corresponding to the axial-vector part of the hadronic currents, 
only two are model-independent. For our convenience we keep all of them in  Eq. (\ref{eq:fiveF}).

The model-dependence in Eqs. (\ref{eq:twomes}) and (\ref{eq:fiveF}) is included in the hadronic form-factors 
($F_V$, $F_S$ expressed in terms of $F_i$, i = 1...5).
For three meson decay modes, the hadronic form-factors calculated within R$\chi$T can be written as
\begin{equation}
F_I = F_I^\chi + F_I^R + F_I^{RR}
\end{equation}
where $F_I^{\chi}$ is the chiral contribution, $F_I^R$ is the one resonance contribution and $F_I^{RR}$ is the double-resonance part. 
The explicit form of the functions $F_i$ for 3$\pi$ and $KK\pi$ modes can be found in  
\cite{Shekhovtsova:2012ra}, Section 2, and in  \cite{GomezDumm:2003ku,Dumm}. 
The corresponding theoretical form-factors are obtained 
in the isospin limit ($m_{\pi}=(2m_{\pi^+}+m_{\pi^0})/3$, $m_K=(m_{K^+}+m_{K^0})/2$), except for the two pion and two kaons modes.

\section{Numerical results. Fit of the three pion  mode to BaBar data}\label{sec:num}

To avoid problems with multi-dimensional integration of the $a_1$-meson propagator, which is rapidly-changing as a function 
of its arguments, we first tabulated the $\Gamma_{a_1}(q^2)$ of 
 Eq. (38) of \cite{Shekhovtsova:2012ra}. 
Then we use linear interpolation to obtain the value of the $a_1$ width at the required $q^2$.

To check the numerical stability of the generator and the multiple numerical integration, the following tests have been done:
\begin{itemize}
\item For every channel one dimensional spectrum $d\Gamma/dq^2$ (for the three meson decay modes) and 
$d\Gamma/ds$ (for the two meson decay modes) produced by the generator has 
been compared with the semi-analytical results. 
For the three meson decay modes the Gauss integration method has been applied to integrate the analytical results for the hadronic currents.
The results agree within statistical errors except for the first and last bins. 
\item The spectrum obtained by the Gauss integration has been compared with the linear interpolated 
spectrum from the neighboring points and demonstrated that the fluctuations due to numerical 
problems of integration are absent;
\item The total rate for every channel obtained from a Monte Carlo run has been compared with the semi-analytical method and required to agree within statistical error.
 The results of comparison are presented in Tab.\ref{tab:bench}.
\end{itemize}
The tests have been performed in the isospin limit  of meson masses,
except for the two pion and two kaons modes. For more details about these tests see \cite{Shekhovtsova:2012ra}. 
Triple Gaussian integration is used for the analytical calculation and double Gaussian integration is used 
for the current calculation that enters the matrix elements of the Monte Carlo generation. 
Thus, a pre-tabulation of the $a_1$ width, $\Gamma_{a_1}(q^2)$, is a convenient tool to significantly increase the generation speed. 

\begin{table*}
\begin{center}
{ \begin{tabular}{|r| c| c| c|} 
\hline
Channel & \multicolumn{3}{c|}{Width, [GeV]} \\
\cline{2-4}
    & PDG  & Equal masses &   Phase space\\  
   &  &  &  with masses \\  
\hline
{$ \pi^-\pi^0 \; \;\; \;$}  &      {($5.778 \pm 0.35\%)\cdot 10^{-13}$} &  {($5.2283 \pm 0.005\%)\cdot10^{-13}$} &  {$(5.2441\pm 0.005\%)\cdot 10^{-13}$} \\ 
{$ K^-\pi^0 \; \;\; \;$}   &       {($9.72\;\pm 3.5\%\;)\cdot 10^{-15}$} &  {($8.3981 \pm 0.005\%)\cdot10^{-15}$} &  {$(8.5810\pm 0.005\%)\cdot 10^{-15}$} \\ 
{$ \pi^-\bar K^0 \; \;\; \;$}   &  {($1.9\;\;\; \pm 5\%\;\;\;)\cdot 10^{-14}$} &  {($1.6798 \pm 0.006\%)\cdot10^{-14}$} &  {$(1.6512\pm 0.006\%)\cdot 10^{-14}$} \\ 
{$ K^-K^0 \; \;\; \;$}     &       {($3.60\; \pm 10\%\;\;)\cdot 10^{-15}$} &  {($2.6502 \pm 0.007\%)\cdot10^{-15}$} &  {$(2.6502 \pm 0.008\%)\cdot 10^{-15}$} \\ 
 {$  \pi^-\pi^-\pi^+$} &           {($2.11\; \pm 0.8\%\;\;)\cdot 10^{-13}$} &  {($ 2.1013\pm 0.016\%)\cdot10^{-13}$} &   {$(2.0800\pm 0.017\%)\cdot 10^{-13}$}  \\ 
{$  \pi^0\pi^0\pi^-$}  &           {($2.10\; \pm 1.2\%\;\;)\cdot 10^{-13}$} &  {($ 2.1013\pm 0.016\%)\cdot10^{-13}$} &  {$(2.1256\pm 0.017\%)\cdot 10^{-13}$}\\ 
 {$  K^-\pi^-K^+$} &   {($3.17\; \pm 4\%\;\;\;)\cdot 10^{-15}$} &  {($3.7379 \pm 0.024\%)\cdot10^{-15}$} &   {$(3.8460\pm 0.024\%)\cdot  10^{-15}$}  \\ 
{$  K^0\pi^-\bar{K^0}$}  &         {($3.9\;\; \pm 24\%\;\;)\cdot 10^{-15}$} &  {($3.7385 \pm 0.024\%)\cdot10^{-15}$} &  {$(3.5917\pm 0.024\%)\cdot 10^{-15}$}\\
{$  K^-\pi^0 K^0$} &               {($3.60\; \pm 12.6\%\;\;)\cdot 10^{-15}$} &  {($2.7367\pm 0.025 \%)\cdot10^{-15}$} &  {$(2.7711 \pm 0.024\%)\cdot 10^{-15}$}\\
\hline
\end{tabular} 
}  
\end{center}
\caption{The $\tau$ decay partial widths. 
The PDG value \cite{Nakamura:2010zzi}, the  2nd column, is compared with numerical results from  \texttt{TAUOLA}
 with the R$\chi$T currents in the 3rd column (obtained in the isospin limit for meson masses) and the 4th column (using physical masses for the pahse space).
%
} \label{tab:bench}
\end{table*}

To obtain the proper kinematic configurations, the differences between neutral and charged pion and kaon masses were taken into account, more specifically, the physical values were chosen
in the phase space generation. 
On the other hand, this choice breaks constraints resulting from isospin symmetry in potentially uncontrolled way.  
This is why we collect the numerical results from the Monte Carlo calculation, shown in Table \ref{tab:bench}, 
where the partial widths from Particle Data Group compilation \cite{Nakamura:2010zzi} are compared 
with our results obtained with isospin-averaged pseudoscalar masses and with the physical ones.
The model parameters, more specifically the masses of the resonances and the coupling constants, were fitted to Aleph data \cite{Barate:1998uf}, requiring 
correct high-energy behavior of the related form factors. For details  see Appendix C in \cite{Shekhovtsova:2012ra}. 
From Table \ref{tab:bench}, one can see that the difference between the TAUOLA results for 
the $\tau$ decay partial width and the PDG ones is $1.5\% -17\%$ depending on the mode.
As expected, the agreement is not good because only minimal attempts to adjusting the model 
parameters have been applied for the comparison with BaBar and Belle data.

\begin{figure*}\label{fig:fit}
\begin{center}
\includegraphics[width = 0.26\textwidth]{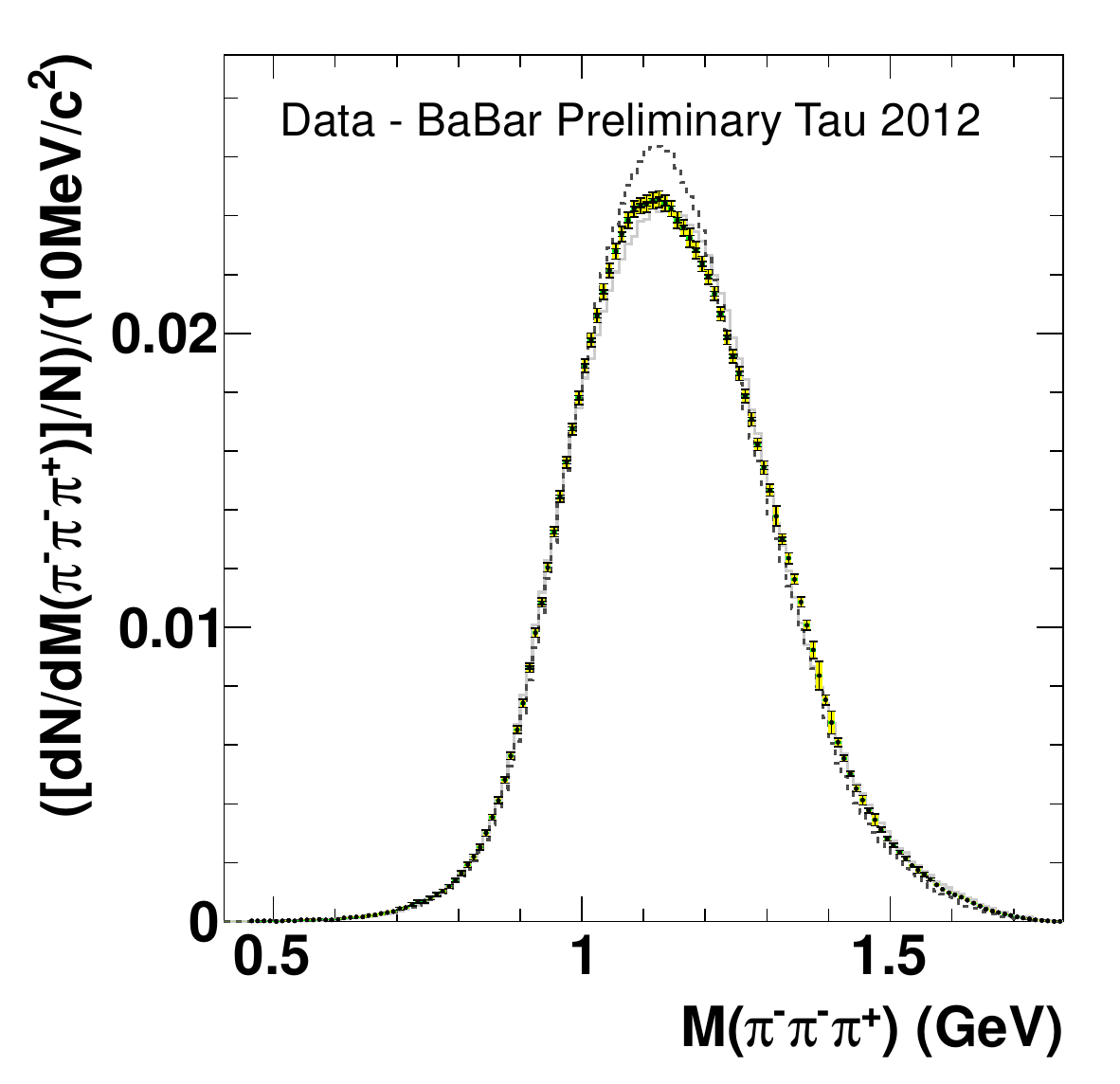}
\hspace{0.5cm}
\includegraphics[width = 0.26\textwidth]{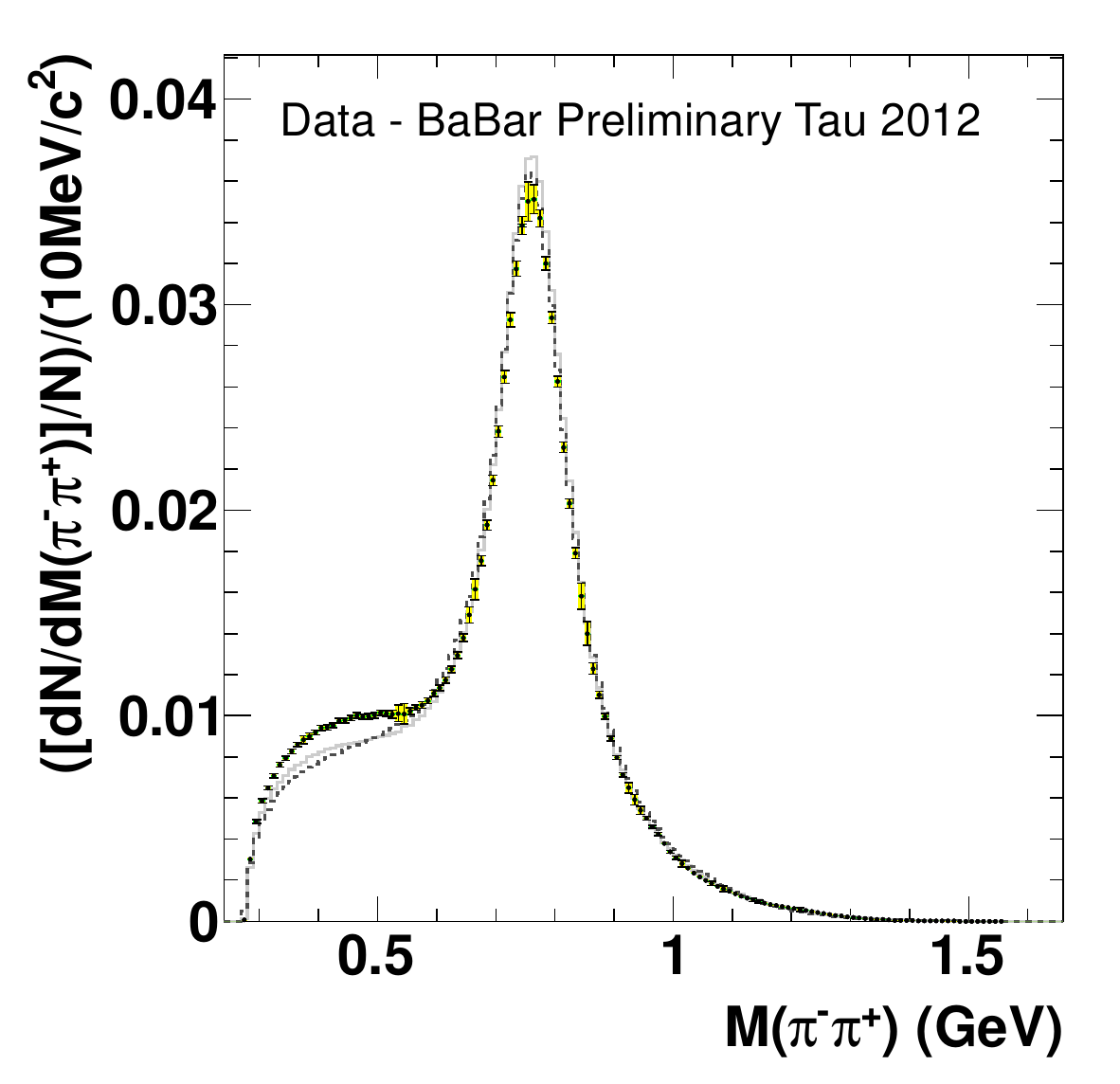}
\hspace{0.5cm}
\includegraphics[width = 0.26\textwidth]{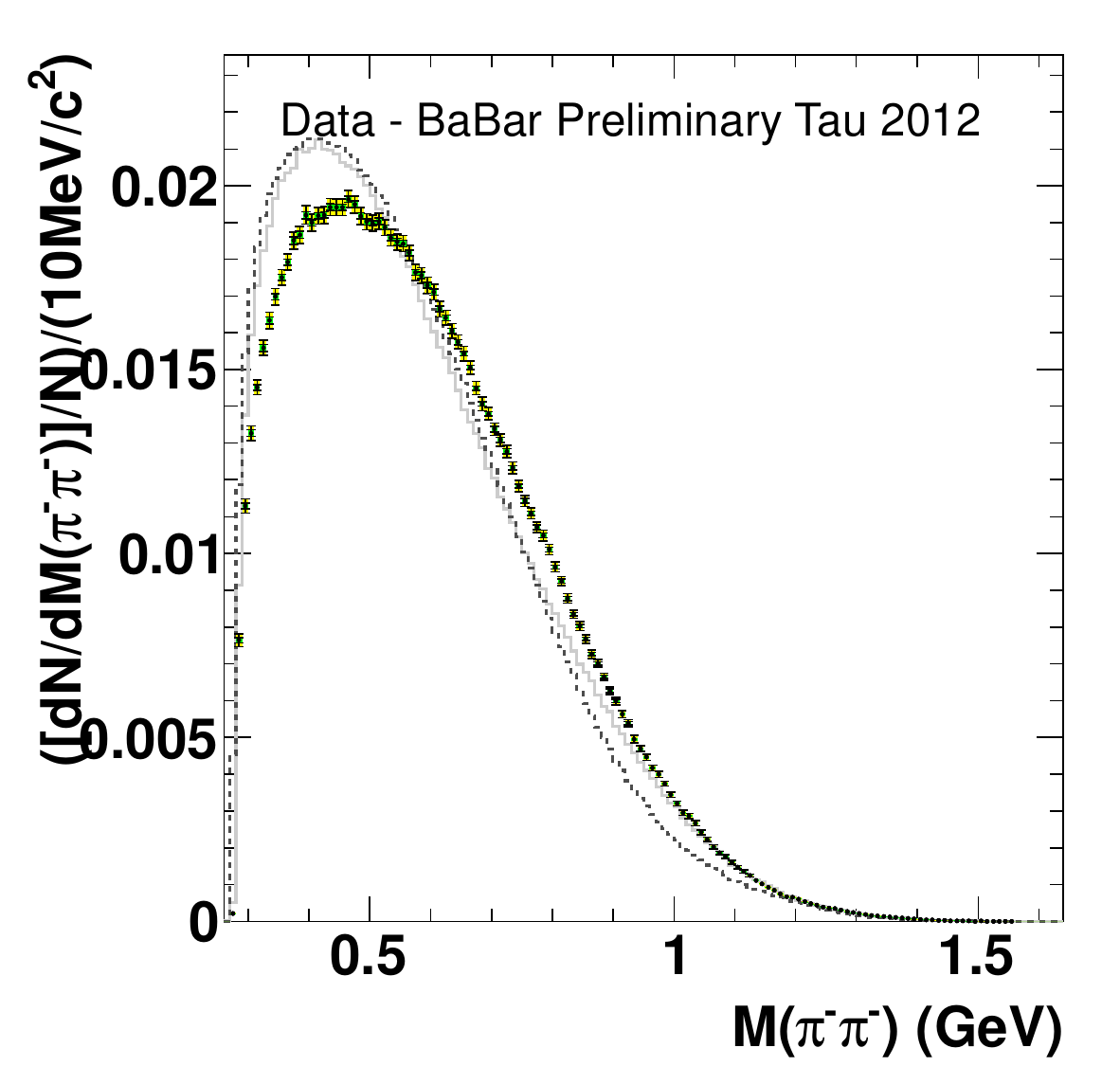}
\caption{Invariant $\pi^+\pi^-\pi^-$ (left) and  
$\pi^+\pi^-$ (center) and $\pi^-\pi^-$ (right) mass distributions.
Lighter grey histograms are for R$\chi$T parametrization, darker grey is for {\texttt CLEO} one, points are Babar data.}
\end{center}
\end{figure*} 

\begin{table*}
\begin{center}
\begin{tabular}{|l|l|l|l|l|l|l|l|}
\hline
  & $M_{\rho'}$& $\Gamma_{\rho'}$& $M_{a_1}$& $F$& $F_V$& $F_{A}$& $\beta_{\rho'}$\\
\hline
Min. & 1.44 & 0.32 & 1.00    & 0.0920 & 0.12 & 0.1 & -0.36\\
\hline
Max & 1.48 & 0.39 & 1.24    & 0.0924 & 0.24 & 0.2 & -0.18 \\
\hline
Default & 1.453 & 0.4 & 1.12    & 0.0924 & 0.18 & 0.149 & -0.25 \\
\hline
Fit , $\chi^2/ndf = 2262.12/132$ & 1.4302 & 0.376061 & 1.21706    & 0.092318 & 0.121938 & 0.11291 & -0.208811\\
\hline
\end{tabular}
\end{center}
\caption{Numerical values of the R$\chi$T parameters fitted to BaBar data for three pion mode \cite{Nugent:2009zz}}
\label{tab:fit}
\end{table*}
Currently, only the differential spectrum of 
the two pion modes \cite{belle} and three pseudoscalar modes
\cite{Nugent:2009zz} are published.  We begin with a fit to the spectrum of the $\pi^+\pi^-\pi^-$ mode. 
The result is presented in Table \ref{tab:fit} and Fig.\ref{fig:fit}.
The fit was done taking into account only the dominant S-wave mechanism (we follow the determination done in \cite{Shibata:2002uv}). 
As suggested in \cite{Shibata:2002uv} the discrepancy in the low mass region could be described adding a contribution 
from a scalar particle, $P$-wave mechanism.
We expect that inclusion of 
the lowest-lying scalar resonance \cite{sigma} 
 will improve the value of $\chi^2$. 
The values in the 5th row of Table \ref{tab:fit}
are only the preliminary results. They do not necessarily correspond to the minimum of $\chi^2/ndf$ of the final fit. 
Work is in progress. Some technical aspects of the fitting strategy is 
given in talk by Z. Was  \cite{was_proceed}.

\section{Tau physcis at  LHC}

 From the perspective of high energy experiments, such as those at LHC, a good understanding of tau leptons properties contributes important ingredients of new physics signatures. 
With the discovery of a new particle around the mass of 
125-126 GeV \cite{higgs:2012}, tau decays are an important decay mode for determining if this is the Standard Model Higgs.
This is especially pertinent since CMS has reported a deficit in the 
number of fermion decays from the new particle relative to the Standard 
Model Higgs Prediction.

At LHC, at the moment, tau decays are only used for identification and are not used to study their dynamic. 
However, the dynamics of tau decays are important for both modeling the decays and therefore the reconstruction, 
identification and for measuring the polarization of tau decays. 
As a result, an upgrade to the TAUOLA
based on the BaBar and Belle measurements of tau
decays is of some interest  for systematic errors evaluation
for LHC measurements.

\section{Conclusion}\label{sec:concl}
The theoretical results for the hadronic currents of two and three pseudoscalar modes, namely, $\pi \pi$, $K\pi$, $K K$, $\pi\pi\pi$ and $K K \pi$, 
 in the framework of R$\chi$T
have been implemented in TAUOLA. These modes, together with the one-meson decay modes, represent more than 88\% of the hadronic width of the tau lepton.   
R$\chi$T is a more controlled QCD-based model than the usually used Breit-Wigner parameterization. 
However, before making conclusion about validity of the model the theoretical results have to 
be confronted with the experimental data. This can be achieved by fitting the model to data. 
 As a consequence of comparisons with the data, some of the theoretical assumptions
may need to be reconsidered too.
Now that the technical work on current installation is complete,
 the work on fitting the data is in progress in collaboration with 
 theoreticians and experimentalists. Therefore, we consider this work as a step towards a theoretically rigorous description of hadronic
tau decay data.

As soon as the theoretical results are able to reproduce Babar and Belle data for the most important for LHC decays into two and three pions
the TAUOLA-RChT version will be installed into LHC environment as it is described in \cite{Davidson:2010rw}.

\section{Acknowledgement}

This research was supported by a Marie Curie Intra European Fellowship within 
the 7th European Community Framework Programme 
(O.S.) and by Alexander von Humboldt Foundation (I.N.),
 by  the Spanish Consolider Ingenio 2010 Programme 
CPAN (CSD2007-00042) and by  MEC (Spain) under Grants FPA2007-60323, FPA2011-23778 (O.S. and P.R.) and FPA2011-25948 (P.R.) and in part by the funds of Polish National Science Centre under decision DEC-2012/04/M/ST2/00240 and DEC-2011/03/B/ST2/00107 (O.S., T.P., Z.W.). 










\end{document}